\documentclass[a4paper,fleqn,usenatbib]{mnras}

\usepackage[T1]{fontenc}
\usepackage{ae,aecompl}

\usepackage{graphicx}	
\usepackage{amsmath}	
\usepackage{amssymb}	
\usepackage{latexsym}
\usepackage{color}

\newcommand{\ltsima}{$\; \buildrel < \over \sim \;$}
\newcommand{\simlt}{\lower.5ex\hbox{\ltsima}}
\newcommand{\gtsima}{$\; \buildrel > \over \sim \;$}
\newcommand{\simgt}{\lower.5ex\hbox{\gtsima}}
\newcommand{\kms}{{\rm\,km\,s^{-1}}}

\newcommand{\pris}{{\textit{Pristine }}}
\newcommand{\umg}{u_{0}\!-\!g_{0}}
\newcommand{\umcahk}{u_{0}\!-\!CaHK_{0}}
\newcommand{\gmr}{g_{0}\!-\!r_{0}}
\newcommand{\teff}{T$_{\rm eff}$}

\def\FeH{{\rm[Fe/H]}}

\title[The Galactic outer halo with BHB stars]{The \emph{Pristine} survey -- VII. A cleaner view of the Galactic outer halo using blue horizontal branch stars}

\author[E. Starkenburg et al.,]{Else Starkenburg$^1$\thanks{E-mail: estarkenburg@aip.de; kyouakim@aip.de}, Kris Youakim$^1$, Nicolas Martin$^{2,3}$, Guillaume Thomas$^4$, 
\newauthor David S. Aguado$^5$, Anke Arentsen$^1$, Raymond G. Carlberg$^6$, Jonay I. Gonz\'{a}lez
\newauthor Hern\'{a}ndez$^{7,8}$, Rodrigo Ibata$^2$, Nicolas Longeard$^2$, Alan W. McConnachie$^4$, 
\newauthor Julio Navarro$^9$, Rub\'en S\'anchez-Janssen$^{10}$, Kim A. Venn$^9$\\
$^1$ Leibniz Institute for Astrophyics Potsdam (AIP), An der Sternwarte 16, D-14482 Potsdam, Germany\\
$^2$ Universit\'e de Strasbourg, CNRS, Observatoire astronomique de Strasbourg, UMR 7550, F-67000 Strasbourg, France\\
$^3$ Max-Planck-Institut f\"{u}r Astronomie, K\"{o}nigstuhl 17, D-69117 Heidelberg, Germany \\
$^4$ NRC Herzberg Astronomy and Astrophysics, 5071 West Saanich Road, Victoria, BC V9E 2E7, Canada \\
$^5$ Institute of Astronomy, University of Cambridge, Madingley Road, Cambridge CB3 0HA, UK\\
$^6$ Department of Astronomy \& Astrophysics, University of Toronto, Toronto, ON M5S 3H4, Canada \\
$^7$ Instituto de Astrof\'{i}sica de Canarias, V\'{i}a L\'{a}ctea, 38205 La Laguna, Tenerife, Spain \\
$^8$ Universidad de La Laguna, Departamento de Astrof\'isica, 38206 La Laguna, Tenerife, Spain \\
$^9$ Dept. of Physics and Astronomy, University of Victoria, P.O. Box 3055, STN CSC, Victoria BC V8W 3P6, Canada \\
$^{10}$ UK Astronomy Technology Centre, Royal Observatory Edinburgh, Blackford Hill, Edinburgh, EH9 3HJ, UK \\
}

\date{Accepted XXX. Received YYY; in original form ZZZ}

\pubyear{2018}

\begin{document}
\label{firstpage}
\pagerange{\pageref{firstpage}--\pageref{lastpage}}
\maketitle

\begin{abstract}
We use the \pris survey $CaHK$ narrow-band photometry, combined with the SDSS $ugr$ photometry, to provide a cleaner sample of blue horizontal branch stars in the Galactic halo out to large distances. We demonstrate a completeness of 91\% and a purity of 93\% with respect to available spectroscopic classifications. We subsequently use our new clean sample of these standard candles to investigate the substructure in the Galactic halo over the \pris footprint. Among other features, this allows for a careful tracing of multiple parts of the Sagittarius stream, providing a measurement independent from other tracers used and reaching larger distances. Moreover, we demonstrate with this clean and complete sample that the halo follows a density profile with a negative power-law slope of 3.5--4.0. As the relatively shallow SDSS $u$-band is the limiting factor in this technique, we foresee large potential for combining Pristine survey photometry with the much deeper $u$-band photometry from the Canada-France-Imaging Survey. 
\end{abstract}

\begin{keywords}
Galaxy: evolution -- Galaxy: formation -- galaxies: dwarf -- Galaxy: abundances -- stars: abundances -- Galaxy: halo
\end{keywords}



\section{Introduction}

In order to measure the 3D structure of our local Universe, we have always been very much dependent on the existence of the various standard candles that, together, form the components of the distance ladder. Standard candles that are suited for the study of the outer regions of our Galaxy include several types of intrinsically variable stars such as RR Lyrae, but also (with lesser precision) M-giants, carbon stars, and blue horizontal branch (BHB) stars. For several decades now, they have been used as important tracers to study the total enclosed mass of the Milky Way galaxy at different distances \citep[e.g.,][]{SommerLarsen89, Norris91} and to study various (kinematical) substructures in the outer Galactic halo \citep[e.g.,][]{Arnold92,Kinman94}.

As the most striking example of substructure in our Milky Way, the long tidal features originating from the Sagittarius dwarf galaxy \citep[first discovered by][]{Ibata94} have received much interest (see \citealt{Law16} for a review). Because the stream wraps around the Galaxy and traces a large range of distances in the Galactic halo, it provides an excellent opportunity to constrain the dark matter potential of our Galaxy and even its shape \citep[e.g.,][]{Ibata01,Helmi04,Johnston05,Law05}. Over the years it has been mapped using a variety of tracers, from M-giants \citep[e.g.,][]{Majewski03,Koposov15}, main-sequence turn-off stars \citep[e.g.,][]{Koposov13,Pila-Diez14,Lokhorst16}, red horizontal branch stars \citep{Shi12}, carbon-rich long-period variables \citep{Huxor15}, RR-Lyrae \citep[e.g.,][]{Hernitschek17,Cohen17,Sesar17b}, K-giants \citep{Liu14,Janesh16}, to BHB stars \citep[e.g.,][]{Fukushima18}. \citet{Belokurov14} use a mixture of different tracers to follow both the leading and trailing tails of the stream and find that their apocenters lie at $R^{L} = 47.8 \pm 0.5$ kpc and $R^{T} = 102.5 \pm 2.5$ kpc respectively. Using RR Lyrae, \citet{Sesar17b} follow the stream beyond this distance and discover a plume of stars 10 kpc beyond the apocenter of the leading arm, and, even more spectacularly, a ``spur'' extending to 130 kpc at the trailing arm apocenter, thus reaching almost 30 kpc beyond the furthest distance at which it had previously been mapped. Whereas the first feature is confirmed using very deep broad-band data selecting BHB stars \citep{Fukushima18}, the second feature has up until now only been detected using RR Lyrae stars and possibly one M-giant member \citep{Li19}. These measurements at large radii are particularly constraining for the modelling of streams in the Galactic halo throughout its kinematical history \citep{Dierickx17,Fardal19}. Better constraints on the different features of the Sagittarius stream clearly enhance our understanding of the formation and evolution of the Milky Way halo.

In this work, we use the \pris survey \citep[][]{Starkenburg17a}, which employs a narrow-band filter around the Ca H \& K absorption features to select BHB stars and, in particular, to distinguish them from the intrinsically fainter population of blue straggler (BS) stars at similar temperatures. The narrow-band filter used by the \pris survey was initially designed for excellent metallicity sensitivity in FGK-type stars and major science cases of the survey include the search for rare extemely metal-poor stars, the study of very metal-poor satellite systems, and the characterisation of metallicity structures in the Milky Way halo \citep[][]{Starkenburg17a,Youakim17,Caffau17,Starkenburg18,Longeard18,Longeard19,Bonifacio19,Aguado19}. However, in this work, we instead make use of the excellent gravity sensitivity of the narrow-band filter for A stars, in particular when it is combined with $u$-band photometry from, for instance, the Sloan Digital Sky Survey (hereafter SDSS; \citealt{york00}). Using broad-band information alone, success rates for separating BHB and BS stars have reported completenesses of 50 to 70 per cent and contamination rates up to 30 per cent \citep[e.g.,][]{Bell10,Vickers12,Fukushima18,Thomas18a}.  Naturally, these success rates drop as fainter stars are investigated and the uncertainties on the photometry increase. \citet{Deason12} report that out of 38 SDSS photometrically selected stars with spectroscopic follow up in the 20 $< g <$ 22 magnitude regime, only 7 bona fide BHB stars were found.  

We demonstrate in this work how the combination of \pris survey and SDSS photometry leads to an unprecedentedly clean and complete sample of BHB stars in Section \ref{sel}. In Section \ref{results} we subsequently focus on characterizing the outer halo using this new BHB sample in the \pris survey footprint. We investigate its spatial coverage in several ways, focusing on the halo profile and quantifying its clumpiness. We additionally use this new sample to map the Sagittarius stream in our footprint, including the ``spur'' feature, which we tentatively trace further out. We conclude our findings and present a future outlook in Section \ref{conc}.

\section{Selection of BHB stars}\label{sel}

\begin{figure*}
\begin{center}
\includegraphics[width=1\linewidth]{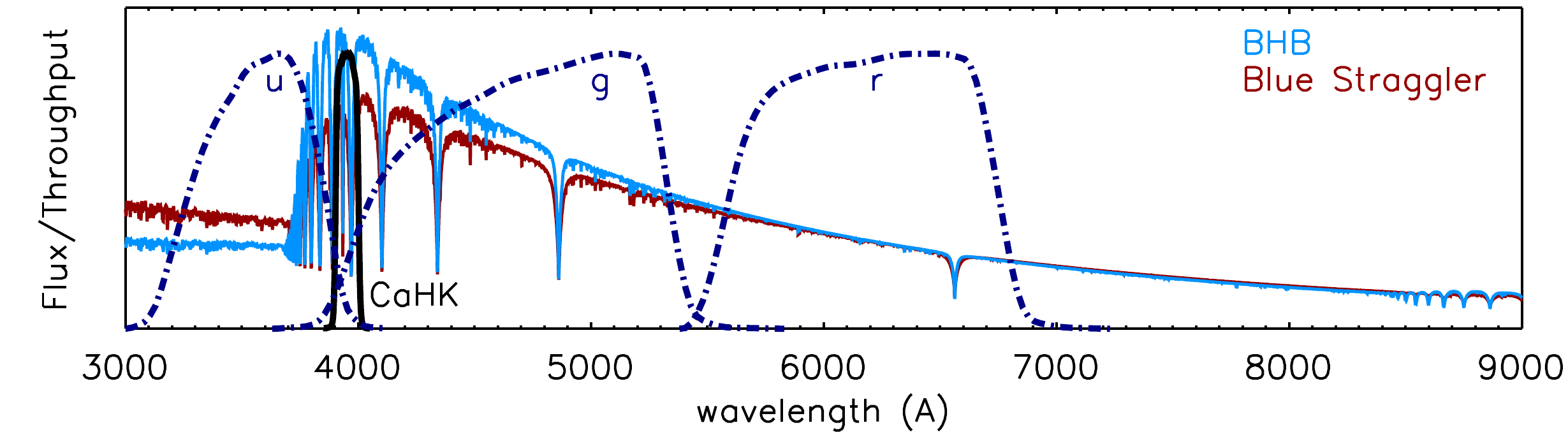}\\
\includegraphics[width=1\linewidth]{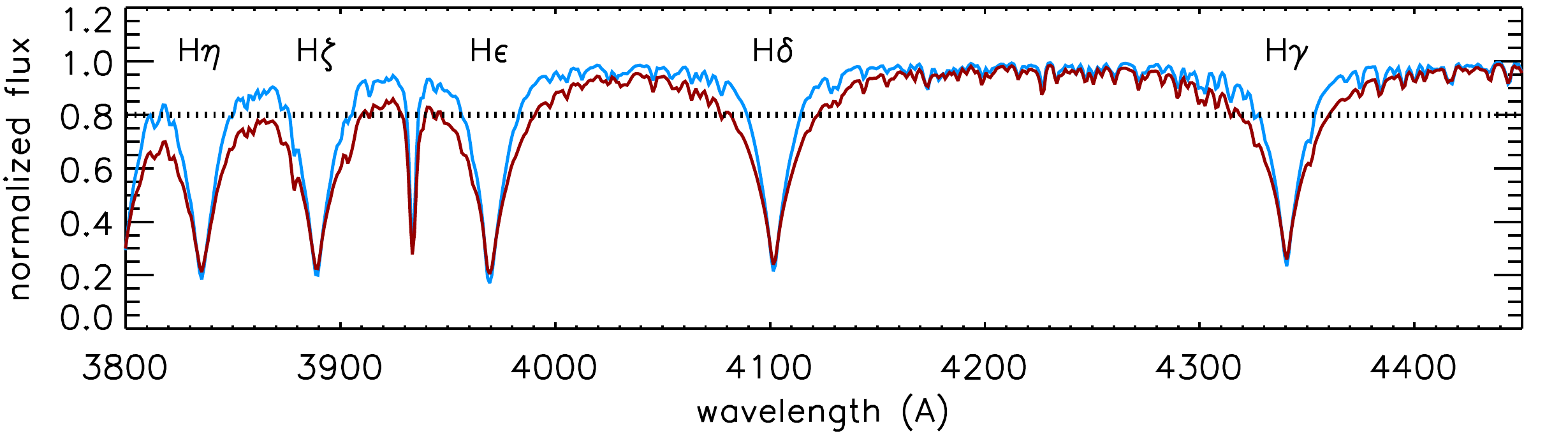}
\caption{Top panel: Synthetic spectra from a library of synthetic stellar spectra by \citet{Munari05} for stars with stellar parameters \teff = 8500 K and $\FeH= -0.5$. One of them has $\log g = 3.0$ (light blue, typical parameters for a BHB star) and the other one $\log g = 5.0$ (dark red, typical for a BS star) overplotted with the throughput curves for the SDSS $ugr$ filters (dash-dotted dark blue lines) and the filter curve for the \textit{Pristine} survey CaHK filter (thicker black line). Bottom panel: The same spectra, but now for a narrower wavelength range such that the Balmer series of hydrogen lines is plainly visible. The horizontal dotted line indicates a level of 0.8 in the normalized spectra where the widths of the hydrogen lines are traditionally measured to determine if the star is a BHB or BS star (see text for details). \label{fig:syntspecBHBBS}}
\end{center}
\end{figure*}

The first selection criterion for BHB stars is a measure of their overall temperature to select A-type stars, as given, for instance, by broad-band SDSS colours. In this region of colour space, there are not many contaminants except for very blue (hot and young) main-sequence stars, which are rarely found in old populations, and blue straggler stars. Careful colour cuts can avoid most contamination from other hot objects such as QSOs \citep{Navarrete18}, white dwarfs \citep{GentileFusillo15}, and hot subdwarfs \citep[][]{Geier18}. In contrast to genuine hot main-sequence stars, blue straggler stars are found in almost all types of populations. In the colour-magnitude diagram, they reside above the main-sequence. The physical difference between them and BHB stars is that they have higher gravities, originating from hydrogen burning stars on the main sequence whereas the BHB stars are giant helium-burning stars. While BHB stars typically have gravities $\log g =$ 2.8--3.75, BS stars show gravities in the range of $\log g = $3.75--5.0 \citep{Vickers12}. Because the latter are intrinsically fainter than BHB stars, it is important to clean any BHB samples of these contaminants before using their standard candle distances to investigate the 3D structure of the Milky Way halo.

This difference in gravity between the two populations in turn leads to differences in their spectra and their broad-band colours. In Figure \ref{fig:syntspecBHBBS}, two synthetic spectra with typical parameters for a BHB star and a BS star of the same temperature are shown, both spectra are taken from a library of synthetic stellar spectra by \citet{Munari05}. From this library, we use the models calculated with the new ODF atmospheres \citep{Castelli03}, no rotation, alpha-enhancement of +0.4, a micro-turbulent velocity of 2 $\kms$, and with 1\AA/pix dispersion. The importance of the $u$-band for distinguishing these two stars stands out clearly, mainly due to the very gravity-sensitive Balmer jump feature which is much steeper for stars of lower gravity \citep[see also][]{Lenz98}. The $g$-band has some gravity sensitivity as well. But at redder wavelengths, where the SDSS $g$-band filter is most efficient, the total flux is dominated by the temperature-sensitive (rather than gravity-sensitive) blackbody curve. Further gravity sensitivity is found to a lesser extent in the Paschen lines in the $z$-band around 8700 \AA. The \textit{Pristine} survey CaHK filter (black solid filter curve) covers a very interesting wavelength range where the difference between the two spectra is at its maximum, but in reversed order to the difference observed in the $u$-band. There is thus reason to expect that a combination of these two filters will provide an excellent diagnostic to discriminate between these two types of stars, even though the \textit{Pristine} survey filter was not specifically designed for this science case.

In addition to photometric measurements at the blue end of the spectrum, BHB and BS stars can be separated by studying their Balmer lines as is illustrated for the same synthetic spectra in the bottom panel of Figure \ref{fig:syntspecBHBBS}. The wings of the Balmer lines are gravity sensitive and the width of these lines (often measured at 80\% --- or alternatively 85\% --- of the total flux level as illustrated by the horizontal dotted line) indicates the gravity of the star \citep[e.g.,][]{Rodgers81,Clewley02,Sirko04,Xue08}. Alternatively, the Balmer lines are fitted by S\'ersic profiles \citep{Sersic63} and the line shape parameters are used to distinguish the different types of stars. Such methods to separate BHB, BS, and main-sequence (MS) stars using the spectra for A-type stars \citep[as pioneered by][]{Pier83} are for instance applied to large spectral libraries of SDSS/SEGUE A-type stars \citep{Yanny00,Xue08,Deason11,Xue11}, or more targeted efforts studying the substructure in the Milky Way outer halo \citep[e.g.,][]{Navarrete18}. Combinations of spectroscopic and photometric measurements are also reported in the literature, such as for instance in \citet{Kinman94,Wilhelm99,Clewley02}. 

In this section we attempt to select a sample that is as clean and pure as possible using either SDSS $ugr$ colours alone (Section \ref{SDSSalone}), or SDSS $ugr$ colours supplemented with the $CaHK$ magnitudes from the \pris survey (Section \ref{SDSSandPris}). We thus focus on characterizing this large and more complete population of BHB stars, for which there is not yet spectroscopic follow-up. Due to their use as standard candles, a clean selection is very important for an accurate characterization of 3D structures in the outer Milky Way halo.  

\subsection{Towards a clean and pure BHB sample using SDSS}\label{SDSSalone}

\begin{figure*}
\begin{center}
\includegraphics[width=1\linewidth]{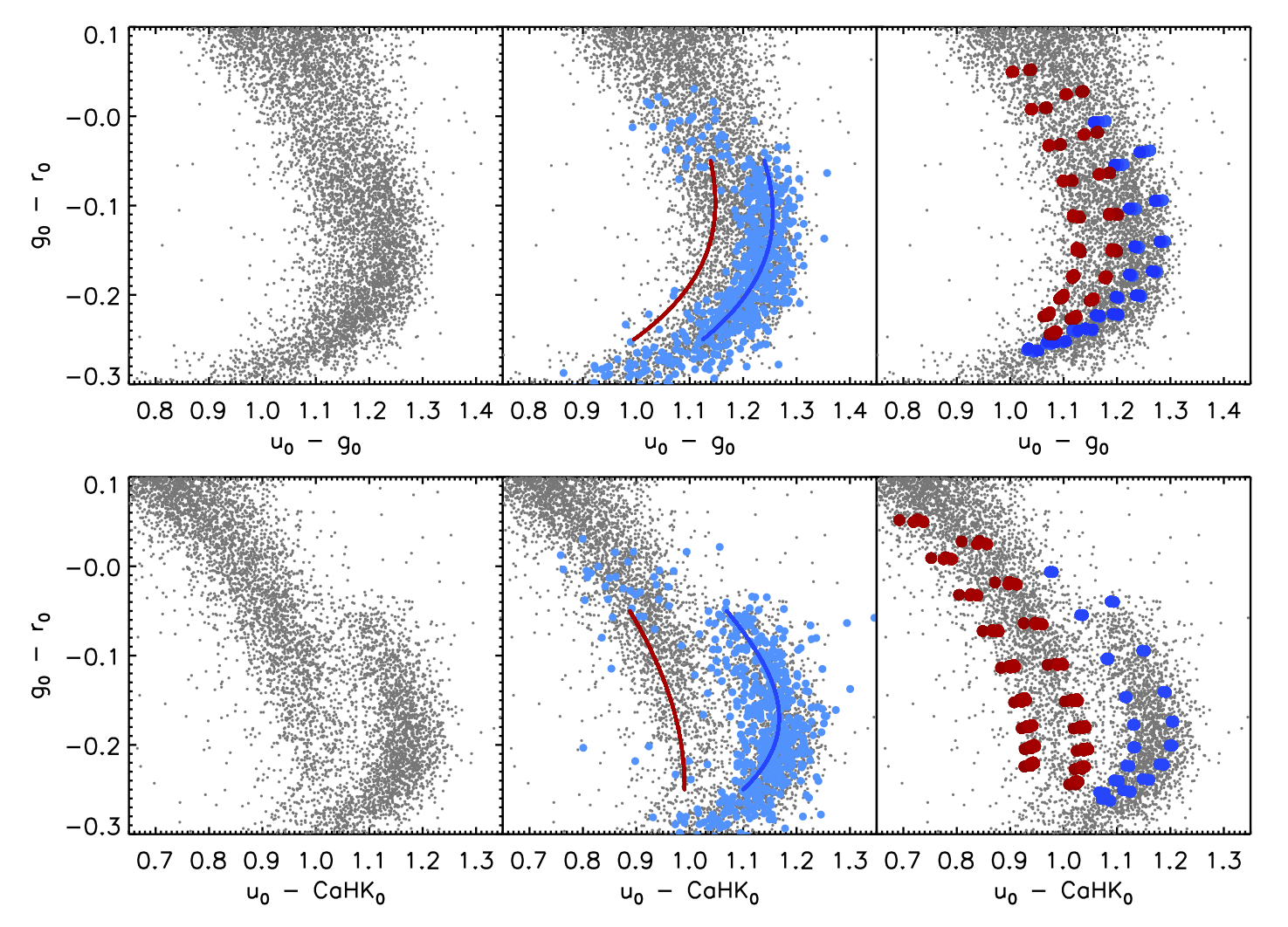}
\caption{Top panels: colour-colour space using SDSS broad-band data for a selected sample of clean photometry for hot stars (see text for details). In the middle, the ridgelines defined in \citet{Deason11} for BHB and BS stars (blue and red lines respectively) are plotted on top of the BHB sample of \citet{Xue11} (light blue filled circles). In the right panel, the results from synthetic colours for spectra from the \citet{Munari05} library with typical BHB (blue points; $-2.5<\FeH<-1.5$, $7500<T_\mathrm{eff}<9500$\,K, $3.0<\log g<3.5$) and BS (red points; $-1.0<\FeH<-0.5$, $7500<T_\mathrm{eff}<9500$\,K, $4.0<\log g<4.5$) stellar parameters are overplotted. Bottom panels: The same stellar sample as in the top panels, but now for a colour-colour space combining the SDSS and the \textit{Pristine} survey photometry. The ridgelines are defined in this work and described in the text. \label{fig:separateBSBHB}}
\end{center}
\end{figure*}

In order to select BHB stars from SDSS colours we closely follow the method outlined by \citet{Deason11}. To start, all objects from the star catalogue of SDSS DR14 \citep{SDSS18} are selected to fulfill the colour cuts:  $0.9 < \umg < 1.4$ and $-0.25 < \gmr < 0.0$, corresponding to the colour-colour space for A-stars. In both these colour cuts the DR14 de-reddened magnitudes are used. As can been seen in Figure 4 of \citet{Deason14}, these broad band cuts ensure that the white dwarf contamination becomes negligible and cut out most QSO contaminants. We subsequently check that the stars are not flagged as variable in Pan-STARRS1 \citep[hereafter PS1,][]{Chambers16}. The variability criteria by \citet{Hernitschek16} provide an independent cleaning by which most QSO sources can be identified (see their Figure 3.). The sample is further cleaned of any star that has not passed the SDSS selection flag for clean photometry, or are flagged in SDSS $u$, $g$, or $r$ photometry separately to be saturated, have de-blending or interpolation problems, are suspicious detections, or are close to the edge of a frame. 

The distribution of these A-type stars in $\umg$ and $\gmr$ colour-colour space is shown in the top-left panel of Figure \ref{fig:separateBSBHB}, where only stars with $u$-band uncertainties less than 0.02 are selected to avoid clutter from low S/N observations. We furthermore restrict ourselves to stars that are also in the \pris survey to facilitate a fairer comparison with the results of the next section. With some effort, two sequences of stars can be distinguished in this panel, although they seem to touch and partly overlap. If the final overlapping sample of spectroscopically confirmed BHB stars from \citet{Xue11} is overplotted (see middle panel), it becomes clear that the rightmost of these sequences corresponds to BHB stars. Any remaining QSOs in the sample are much more likely to contaminate the BS sample rather than the BHB sample, as they predominantly occupy the regions of lower $u-g$ \citep{Deason14}.

\citet{Deason11} have defined two ridgelines to separate the two populations using their classifications from stellar spectra. We here adopt the same ridgelines, but we shift them by 0.04 magnitudes in the $u$ band, since there have been some  shifts in the calibration of the de-reddened magnitudes in the various SDSS data releases \citep{Finkbeiner16} and this is the mean shift for these stars between DR8 and DR14. The lines are overplotted in blue and red, respectively, in the middle top panel of Figure \ref{fig:separateBSBHB}, and described by the following equations: 

\begin{equation}
\begin{split}
	(u_{0}-g_{0})^{0}_{\rm{BHB}} = & 1.167 - 0.775 (g_{0}-r_{0}) - 1.934 (g_{0}-r_{0})^{2} \\
	& + 9.936 (g_{0}-r_{0})^{3} + 0.04,
\end{split}
\end{equation}

\begin{equation}
\begin{split}
	(u_{0}-g_{0})^{0}_{\rm{BS}} = & 1.078 - 0.489 (g_{0}-r_{0}) - 0.556 (g_{0}-r_{0})^{2} \\
	& + 13.444 (g_{0}-r_{0})^{3} + 0.04
\end{split}
\end{equation}

Finally, in the top-right panel of Figure \ref{fig:separateBSBHB}, the values for synthetic models from the Munari stellar library \citep[][using the same sets of synthetic spectra as illustrated in Figure \ref{fig:syntspecBHBBS}]{Munari05} are overplotted, as integrated under the SDSS filter curves. For the synthetic spectra, a shift of 0.04 mag was needed in the $u$-band to correct for the offset between the SDSS $u$ band and its AB magnitude as integrated under the filter curve\footnote{see https://www.sdss.org/dr14/algorithms/fluxcal/}. As in the previous panel, we have additionally added another 0.04 shift since this is the mean shift between DR8 and DR14 photometry in the $u$-band for these stars. Both sets of models are run with effective temperatures ranging from 7500 to 9500 K. Furthermore, the synthetic BHB stars (shown in blue) are selected to have $3.0 \le \log g \le 3.5$ and $-2.5 \le $[Fe/H]$ \le -1.5$, in accord with what is typical for this population \citep{Xue08}. The blue straggler synthetic spectra are selected to have $4.0 \le \log g \le 4.5$ and $-1.0 \le $[Fe/H]$ \le -0.5$. Indeed, the synthetic models with these parameters seem to fit the observed locii of the sample well.

To calculate the probability that a star is either a BHB or a BS star, we again follow the method presented by \citet{Deason11} and, given their measured standard deviations for the BHB and BS populations ($\sigma_{\rm BHB,0}(\umg)=0.04$ and $\sigma_{\rm BS,0}(\umg)=0.045$), we use 

\begin{eqnarray}
\label{eq:prob}
p(ugr~|~{\rm BHB})\propto\mathrm{exp}\left(-\frac{\left[(\umg)-(\umg)_{\rm BHB}^0\right]^2}{2\sigma_{\rm BHB}^2}\right), \notag\\ 
p(ugr~|~{\rm BS})\propto\mathrm{exp}\left(-\frac{\left[(\umg)-(\umg)_{\rm BS}^0\right]^2}{2\sigma_{\rm BS}^2}\right).
\end{eqnarray}

\noindent

Here, 
\begin{eqnarray}
\label{eq:unc}
\sigma_{\rm BHB} = \sqrt{\sigma_{\rm BHB,0}^{2} + \sigma_{\rm BHB,(\umg)}^{2}} \notag\\
 \rm{and}\  \sigma_{\rm BS} = \sqrt{\sigma_{\rm BS,0}^{2} + \sigma_{\rm BS,(\umg)}^{2}},
\end{eqnarray}

\noindent
thus folding in both the intrinsic width of the populations and the uncertainty on the colour measurement. The colour-based posterior probabilities of class membership are then described as:

\begin{eqnarray}
\label{eq:prob2}
P({\rm BHB}~|~ugr)=\frac{p(ugr~|~{\rm BHB})~N_{\rm BHB}}{p(ugr~|~{\rm BHB})~N_{\rm BHB}+p(ugr~|~{\rm BS})~N_{\rm BS}}, \notag\\ 
P({\rm BS}~|~ugr)=\frac{p(ugr~|~{\rm BS})~N_{\rm BS}}{p(ugr~|~{\rm BHB})~N_{\rm BHB}+p(ugr~|~{\rm BS})~N_{\rm BS}}.
\end{eqnarray}

\noindent
Where the total numbers of stars $N_{\rm BHB}$ and $N_{\rm BS}$ in a given colour range are found iteratively and are described in Table~1 of \citet{Deason11}.

\subsection{Towards a clean and pure BHB sample using SDSS and \textit{Pristine}}\label{SDSSandPris}

Here we repeat the selection steps as outlined in the last section. However, instead of using $u_{0}-g_{0}$ and $g_{0}-r_{0}$ as our colour-colour selection space, we replace the former with $u_{0} - CaHK_{0}$; thus taking into account the \textit{Pristine} narrow-band filter. This is illustrated in the bottom panels of Figure \ref{fig:separateBSBHB}. The selected stars, shown here as grey points, are identical to the stars selected for the upper panels. It is clear that the combination of $u$-band and $CaHK$-band photometry allows for a much cleaner selection of BS and BHB stars, as expected from the synthetic spectra in Figure \ref{fig:syntspecBHBBS}. Now, clearly, two different sequences can be distinguished. When the spectroscopically confirmed BHB stars from \citet{Xue11} are overplotted (in the bottom-middle panel), it is clear that the BHB stars are on the sequence on the right. For the coolest stars, with the highest $g_{0}-r_{0}$ values, we do see that some spectroscopically classified BHB stars fall onto the bluer sequence. However, taking into account that the spectroscopic classification in this stellar parameter regime becomes more difficult \citep[see Figure 5 of][]{Xue08}, we actually take this as likely evidence that these stars are spectroscopically misclassified. We note that, independently, a similar conclusion is reached by \citet{Lancaster19}, who remove stars with $\umg < 1.15$ and $\gmr > -0.07$ from the sample based on their high tangential velocities if a BHB star distance is adopted (indicating they are probably misclassified BS stars). For the remainder of this work, we restrict ourselves to $g_{0}-r_{0} < -0.05$.    

As in the previous section, we find that the synthetic colours derived from synthetic spectra integrated under the filter curves do a good job in following the two sequences. We therefore let the synthetic predictions guide the definition of our ridgelines in this new colour-colour space, by fitting the synthetically predicted points in the right bottom panel with polyomials. The ridgelines we obtain are shown in the bottom middle panel and are described by

\begin{equation}
\begin{split}
	(u_{0}-CaHK_{0})_{\rm{BHB}} =\ & 0.997 - 1.465 (g_{0}-r_{0}) + 0.411 (g_{0}-r_{0})^{2} \\
	& + 18.531 (g_{0}-r_{0})^{3},
\end{split}
\end{equation}

\begin{equation}
\begin{split}
	(u_{0}-CaHK_{0})_{\rm{BS}} =\ & 0.832 - 1.222 (g_{0}-r_{0}) - 2.094 (g_{0}-r_{0})^{2} \\
	& + 1.046 (g_{0}-r_{0})^{3}.
\end{split}
\end{equation}

We remove from the sample any objects that are beyond 3$\sigma$ from any of these two ridgelines (taking into account the dispersion and photometric uncertainties as outlined in Equation \ref{eq:unc}) to remove spurious objects and contamination. This only removes $\sim$1\% of all objects. Subsequently, we again follow the procedure as described in Equations \ref{eq:prob} until \ref{eq:prob2}, with the difference that we replace $\umg$ with $\umcahk$ and that we set both N$_{\rm BHB}$ and N$_{\rm BS}$ to 1, since the sequences are so well separated that no strong prior on the ratios of BS and BHB stars as a function of colour is needed. This allows us to be more agnostic about the fraction of BS and BHB stars as a function of colour and distance.

\subsection{Comparison between the two samples}\label{sec:comp}

In this section we compare the performance of both the samples defined above in both purity and completeness of their BHB selection. For this purpose, we use the spectroscopically defined samples of \citet{Xue08}, which includes both BS and BHB stars, as well as genuine MS stars, that are spectroscopically classified. This sample covers the magnitude range $14.0< g<19.2$. As mentioned above, a cut is imposed on $\gmr$ where we trust the spectroscopic analysis at $< -0.05$. Furthermore, the following quality cuts are applied to both samples to remove sources with bad photometry in either SDSS or the \textit{Pristine} survey: $0.7 < \umcahk < 1.3$, CASU photometry flag for \pris photometry equals -1 (object has a stellar point-like point spread function), the star shows no sign of variability in PS1 photometry, its SDSS photometry is clean, and the uncertainty on $CaHK < 0.1$. The last criterion only removes very few stars, as the overlap with the spectroscopic sample of \citet{Xue08} contains mostly relatively bright stars and the mean uncertainty on $CaHK$ within this sample is 0.013. We also note that in this sample overlapping with spectroscopic observations the number of BS/MS stars is only slightly larger than the number of BHB stars. The natural expectation is that this ratio will vary with height above the Galactic plane (in this sample, the Galactic latitude is always greater than 20 degrees).

\begin{figure}
\begin{center}
\includegraphics[width=1\linewidth]{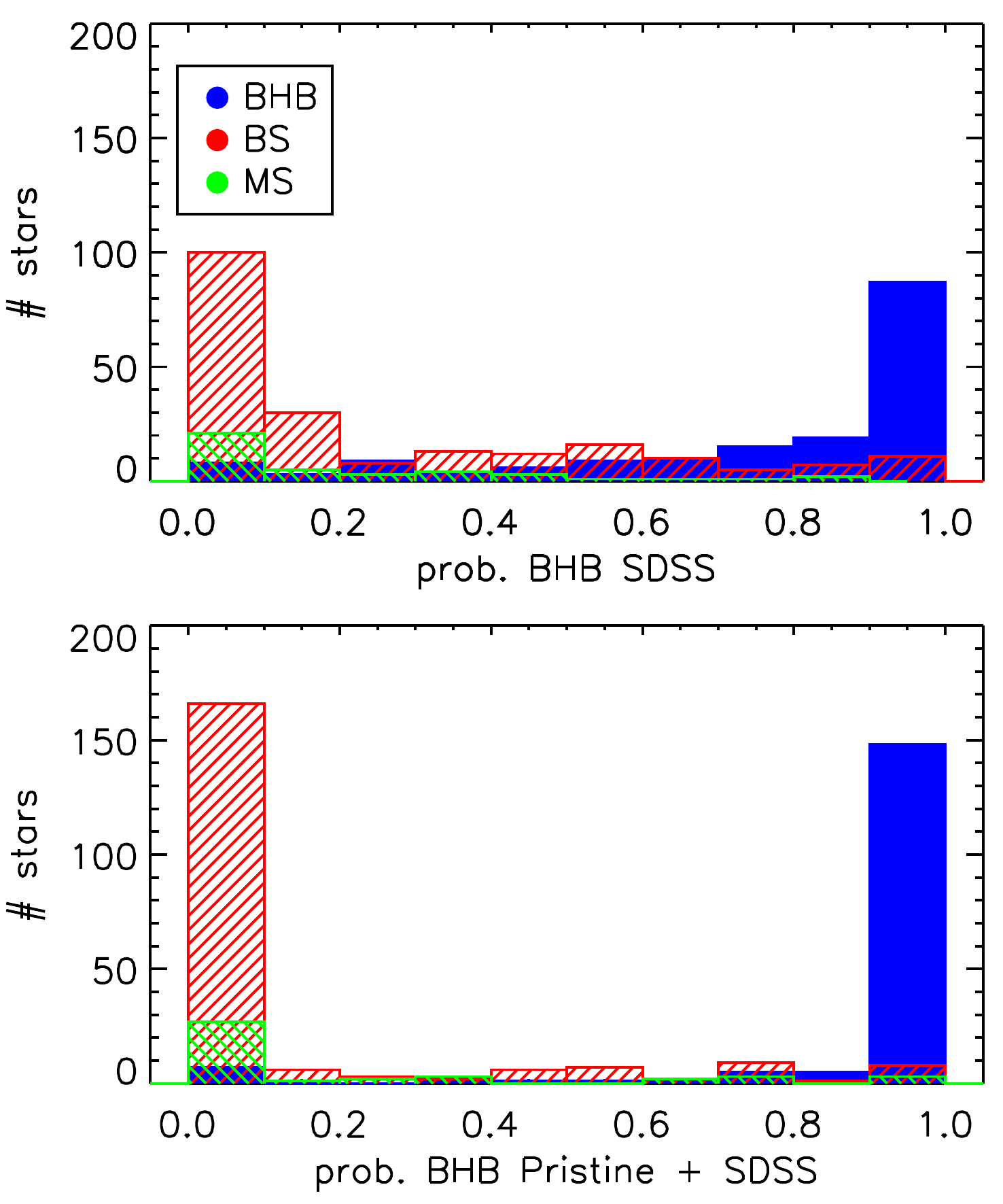}
\caption{The probability that stars from the \citet{Xue08} spectroscopically followed up sample are BHB stars according to the methods presented in Sections \ref{SDSSalone} and \ref{SDSSandPris}. The stars from \citet{Xue08} are held against the same quality criteria (see text for details) and are separated in BHB (blue filled histogram), BS (red histogram), and MS (green histogram) stars according to the \citet{Xue08} classification. It is clear that the addition of $CaHK$ photometry adds to both the purity and completeness of a photometrically selected BHB sample. \label{fig:success}}
\end{center}
\end{figure}

Figure \ref{fig:success} shows for both samples the resulting distribution of the \citet{Xue08} sample as a function of the derived $P({\rm BHB}~|~ugr)$ or $P({\rm BHB}~|~ugrCaHK)$. This figure confirms and quantifies the qualitative result from Figure \ref{fig:separateBSBHB}. With the addition of the \pris\ $CaHK$ photometry, the selection of BHB stars becomes both purer and more complete. Were one to select stars with $P({\rm BHB}) > 0.8$ in both samples, the purity of BHB stars in the SDSS sample is 84\% (16\% is still contamination), while the completeness is 63\% (37\% of the spectroscopically confirmed BHB stars are not selected). These numbers are very similar, or even a bit better, to those quoted in the literature \citep{Sirko04, Bell10, Vickers12}, also when the $z$ band is used, instead or in addition to the use of the $u$ band for classification \citep[see respectively][]{Fukushima18, Thomas18a}. 

When adding the \pris $CaHK$ magnitudes however, as shown in the bottom panel of Figure \ref{fig:success}, the purity rises to 93\% and the completeness rises spectacularly to 91\%.

\subsection{Completeness and purity as a function of magnitude}

\begin{figure}
\begin{center}
\includegraphics[width=1\linewidth]{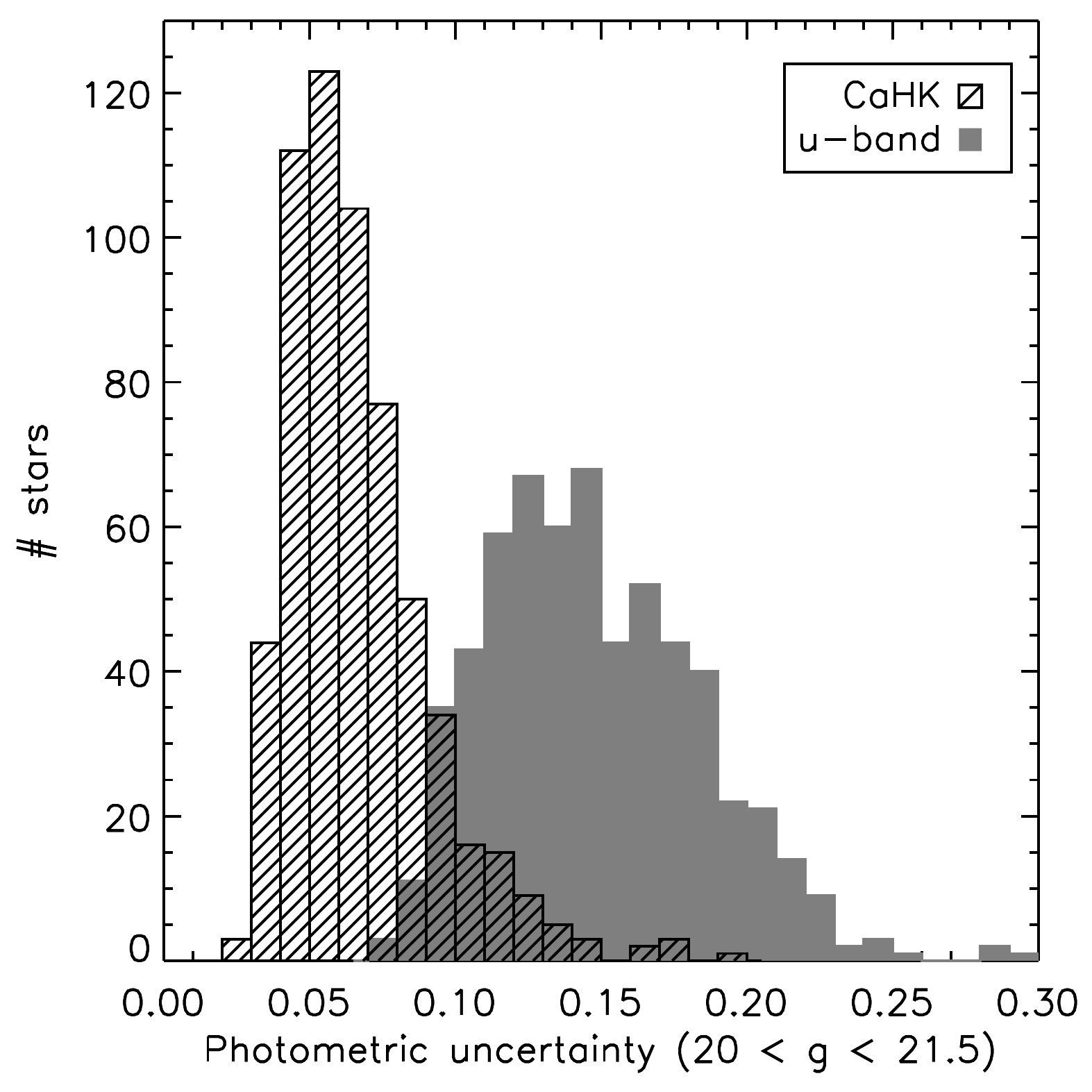}
\caption{Photometric uncertainties in $u$ and $CaHK$ measurements respectively, for A stars with $g$-magnitudes between 20 and 21.5. These would probe distances of 66--170 kpc when classified as BHB stars (depending also on their colour).\label{fig:unc}}
\end{center}
\end{figure}

\begin{figure*}
\begin{center}
\includegraphics[width=1\linewidth]{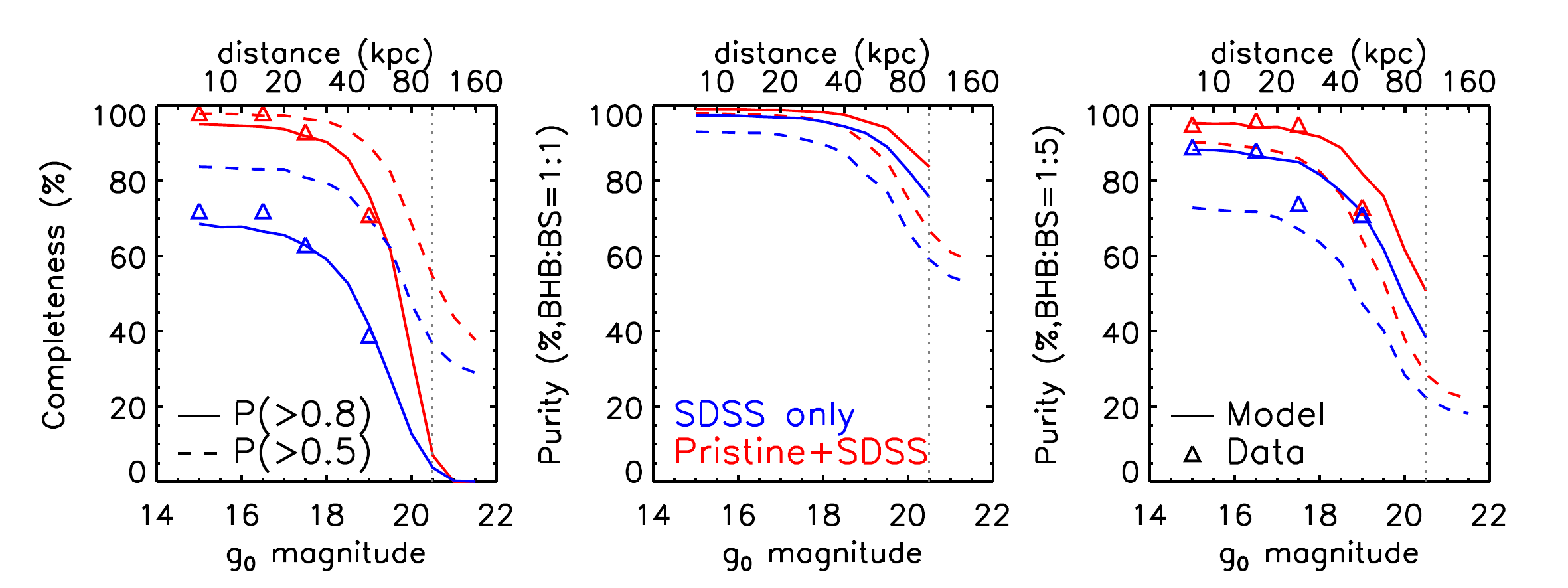}
\caption{Completeness and purity of the BHB sample evaluated as a function of magnitude for the SDSS only and \pris + SDSS methods (blue and red, respectively). The triangles show the results of the overlapping spectroscopic sample of \citet{Xue08}, when binned in magnitude (the magnitude shown represents the middle of the bin), using a probability cut-off value $> 0.8$. The lines show a Monte Carlo model of performance based on the photometric uncertainties. Full lines use a probability cut-off value $> 0.8$, dashed lines $> 0.5$. The two rightmost panels both show results for the purity of the sample, but based on a different underlying modelled ratio of BHB to BS stars, either 1:1 (middle panel), or 1:5 (right panel). The top x axes show a distance scale for a BHB star with ($\gmr$) = -0.15 at that apparent magnitude. A vertical dotted grey line indicates a distance of 100 kpc. }\label{fig:evmag}
\end{center}
\end{figure*}

Hampering a better understanding of the outermost parts of the Galactic halo are the larger photometric uncertainties at fainter magnitudes. BHB stars of 21.5 in $g$-band magnitude are required to enable a mapping of the halo to distances of $130 - 170$ kpc. A clean sample of such stars provides not only a detailed 3D map of the outer halo, but also serves as an excellent sample for spectroscopic follow-up, as such enabling careful kinematic studies of the outer (dark matter) halo. The distance in $\umg$ between the BHB and BS star ridgelines is only 0.12 mag. This difference increases to $\sim$ 0.2 mag in the $\umcahk$ space we use in this work. For stars with $21 < g_{0} < 21.5$, the mean $CaHK$ uncertainty is 0.11 mag and the mean uncertainty on SDSS $u$-band photometry is 0.21 mag (as illustrated in Figure \ref{fig:unc}), meaning that photometric uncertainties will play an important role when attempting to separate both populations. 

In Figure \ref{fig:evmag}, we investigate the results of purity and completeness as a function of magnitude in two ways. Firstly, we investigate the magnitude dependence in the overlapping spectroscopic sample of \citet{Xue08}. Secondly, we Monte Carlo simulate mock datasets with uncertainties typical for that magnitude and subsequently apply our techniques.

For the spectroscopic overlap sample, the numbers of bona fide BHB stars drop significantly at $g_{0} = 18$. Only 31 BHB stars are available between $18 < g_{0} < 20$, making a critical assessment of the performance as function of magnitude difficult. Nevertheless, using this approach, this faintest magnitude bin gets 71\% (53\%) purity, and only 39\% (65\%) completeness by using SDSS only and probability cut-off values $> 0.8$ ($> 0.5$). For \pris + SDSS, the purity remains similar, but the completeness is much better at 71\% (84\%). These values are shown as triangle symbols in Figure \ref{fig:evmag}. It can also be seen in this figure that for stars brighter than $g_{0} = 18$ the \pris + SDSS method shows hardly any decline of success with magnitude. 

Due to the small number of stars in the overlapping spectroscopic sample at faint magnitudes, we also Monte Carlo simulate mock datasets to estimate completeness, which produces a more robust representation of the effect of the photometric uncertainties on the results. Although this approach does of course not include systematic effects, it can be seen from the comparison of the modelling (lines) and data (triangles) in Figure  \ref{fig:evmag} in the magnitude ranges where the two overlap that the model provides a reasonable description of the data. Predictions for performance of completeness are more straightforward to make than for purity, as measurements of purity also depend on the true number ratio between BHB and BS stars (it includes a measure of mis-classified BS stars). Therefore, purity predictions are shown in two panels in Figure \ref{fig:evmag}. In the middle panel, the ratio between BHB and BS stars is modeled as 1:1, in the right panel as 1:5. The latter is more in concordance with the ratio at the faintest magnitude bin in the spectroscopic data set, but this is not necessarily a representation of the true value due to the non-trivial way in which BHB star candidates enter the SDSS selection function \citep[see e.g.,][]{Sirko04}. 

\begin{figure}
\begin{center}
\includegraphics[width=\linewidth, angle=0]{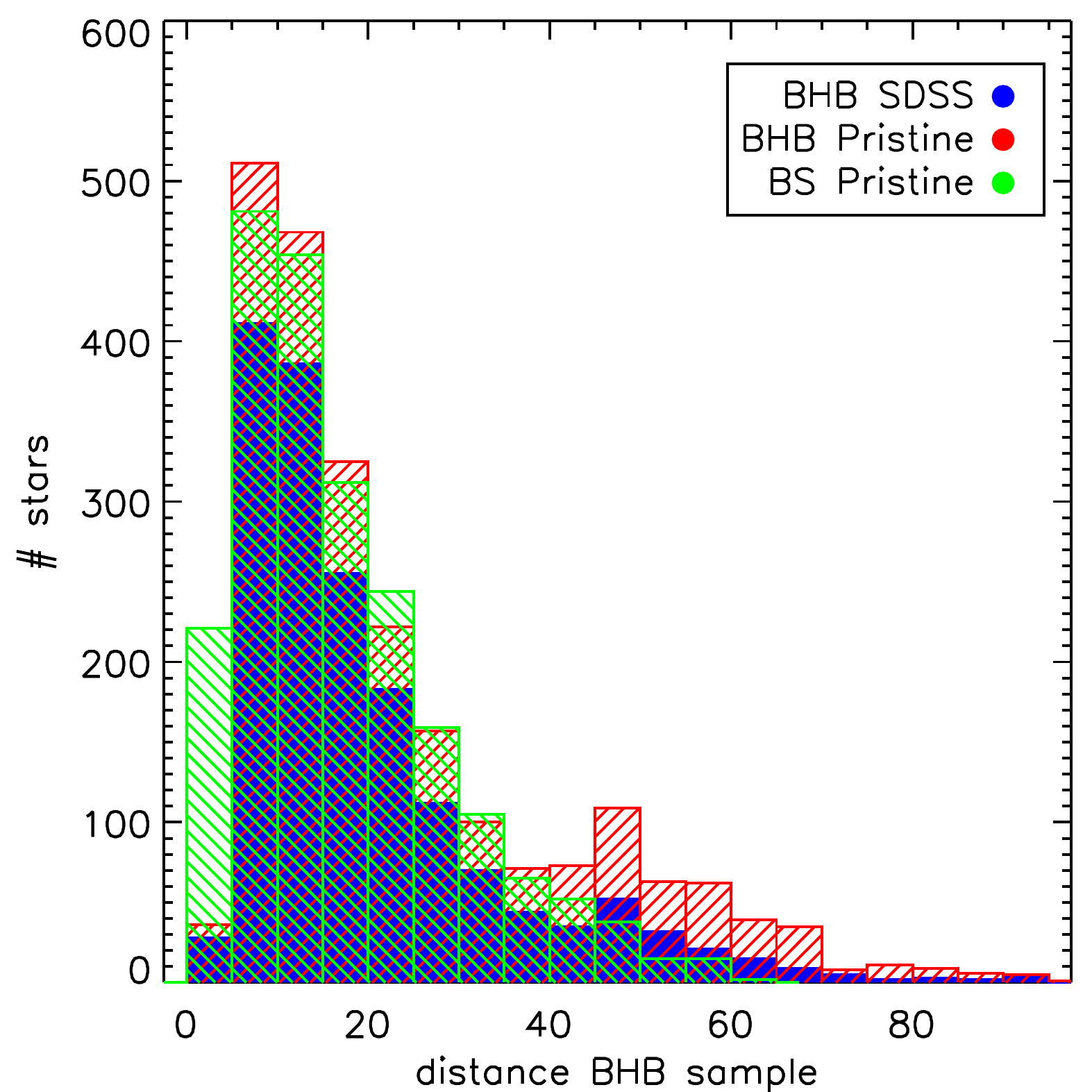}
\caption{Histogram of distances derived for the BHB population as identified in Section \ref{SDSSalone} from SDSS photometry alone (blue solid histogram) and the BHB population as well as the BS population from combining \pris and SDSS photometry as described in Section \ref{SDSSandPris} (red 45$^\circ$ and green 135$^\circ$ slanted histograms respectively). Targets are selected to have clean photometry. The relations to derive absolute magnitudes for these populations are given in Equations \ref{eq:distBHB} and \ref{eq:distBS}. \label{fig:disthist}}
\end{center}
\end{figure}

\begin{figure*}
\begin{center}
\includegraphics[width=\linewidth, angle=0]{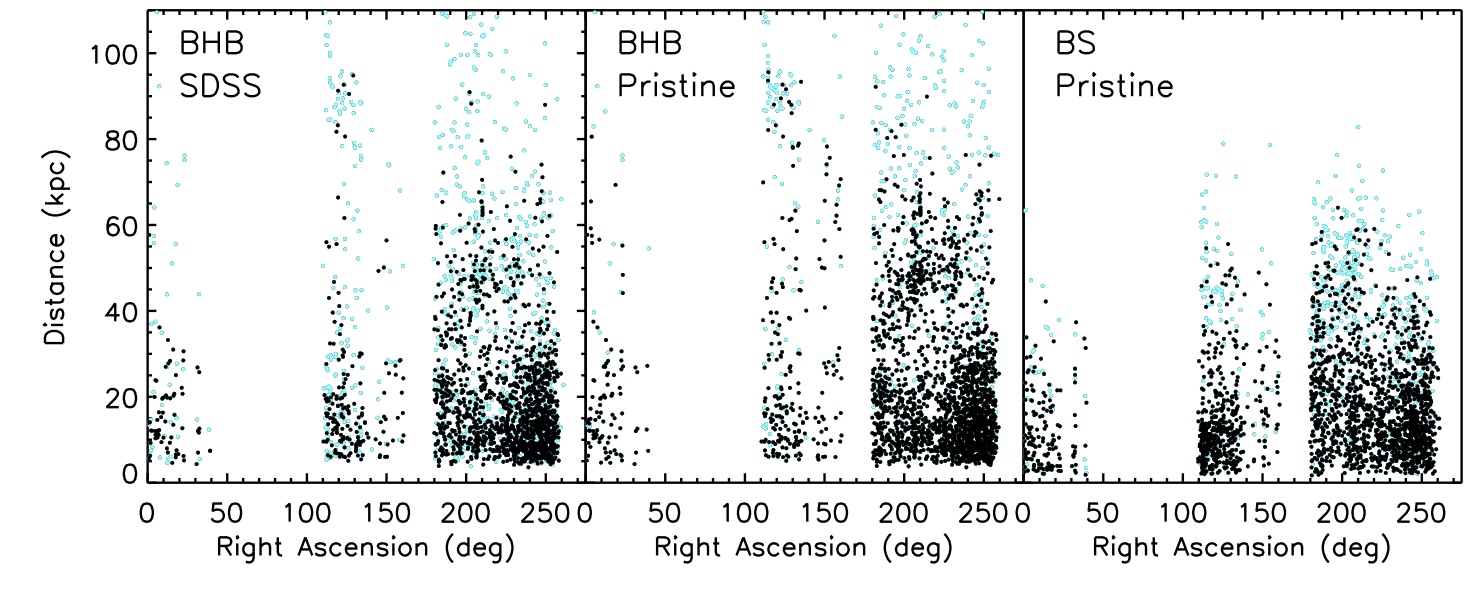}
\caption{Right ascension versus distance maps for the three different populations shown in Figure \ref{fig:disthist}. Stars with a probability $> 0.8$ per class are shown in black, while stars with a probability between 0.5 and 0.8 appear as cyan points. \label{fig:skydist}}
\end{center}
\end{figure*}

\begin{figure*}
\begin{center}
\includegraphics[width=1\linewidth]{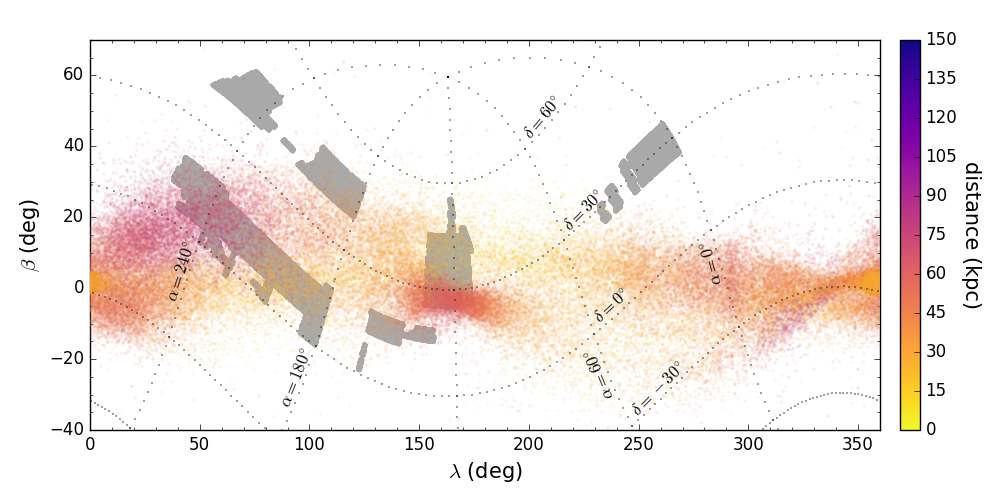}
\caption{Current footprint of the \pris survey (grey filled contours) used in this work in the coordinate system of the orbital plane of the Sagittarius stream (see, e.g., \citealt{Belokurov14} for its definition). Overplotted is the Sagittarius stream modelled by \citet{Law10}, colour-coded by predicted distance to the various stream features. \label{fig:Sagfootprint}}
\end{center}
\end{figure*}

\begin{figure}
\begin{center}
\includegraphics[width=1\linewidth]{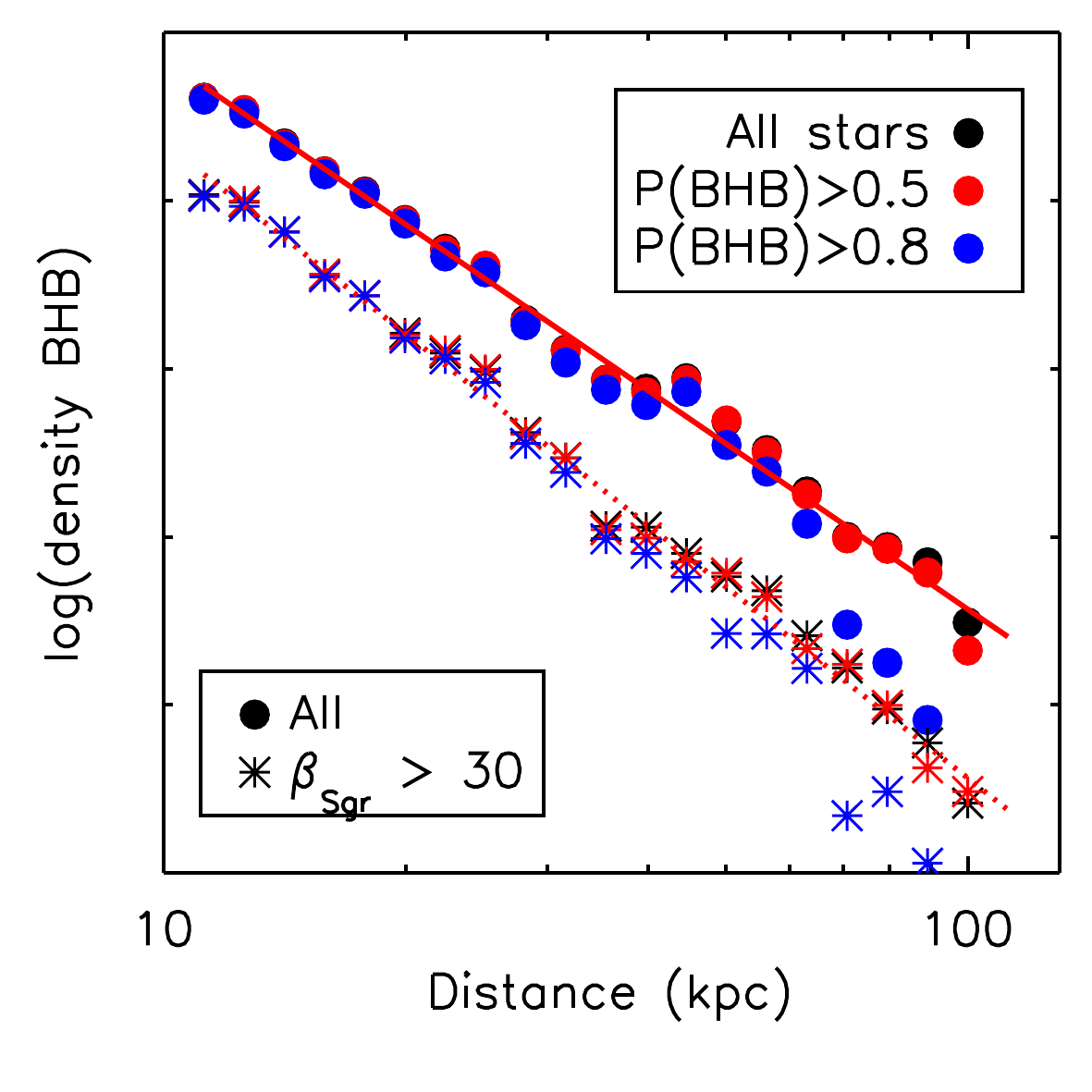}
\caption{Distance profile (in arbitrary units) of BHB stars selected in three different ways: either by taking all stars with a probability that they are a BHB star higher than 0.5 (red filled circles), or 0.8 (blue filled circles), or by taking all stars that pass the quality cuts and multiplying them with the probability they represent BHB stars (black filled circles). A fit to the red symbols is illustrated and has a negative power-law slope of $3.5\pm0.1$ (for the full sample), or $4.0\pm0.1$ for the sample at least 30 degrees off the Sagittarius orbital plane. These profile fits are shown by the full and dotted red lines respectively. \label{fig:profile}}
\end{center}
\end{figure}

On the top x-axes of Figure \ref{fig:evmag} we show the distance corresponding to the magnitude for a BHB star with ($\gmr$) = $-0.15$. The completeness of the P$>0.8$ sample very steeply declines at $g_{0}$-magnitudes beyond 19 and approaches zero at $g_{0} = 20.5$ (corresponding to 100 kpc, indicated by the dotted grey line in Figure \ref{fig:evmag}). Basically, the uncertainties become so large that the method can not distinguish BHB stars with such high probability anymore. The P$>0.5$ sample, on the other hand, shows a completeness of 50\% still at $g_{0} = 21$, but at the cost of a lower purity, the precise value of which will depend strongly on the true number ratio of BHB and BS stars at those magnitudes.

\subsection{Future outlook}
As illustrated in Figure \ref{fig:unc}, the crucial uncertainty in our method is not the $CaHK$ magnitude, but instead the rather shallow SDSS $u$-band photometry (at $g$ = 20 and beyond, $u$-band uncertainties are significantly larger than Pristine photometry uncertainties). Excitingly, the currently on-going Canada-France Imaging Survey \citep[CFIS,][]{Ibata17a,Ibata17b} will provide more accurate $u$-band photometry in coming years. CFIS is $\sim$3 magnitudes deeper than SDSS. CFIS uncertainties in the $u$-band are only 0.01-0.03 mag on average \citep{Ibata17a} in the same magnitude range. The overlap between the \pris and CFIS footprints is a few hundred square-degrees at the moment, but it is quickly increasing and will reach 100\% by the end of the survey. CFIS photometry has already been used to select BHB stars very far out in the Galactic halo \citep{Thomas18a}. With the addition of the \pris photometry this will provide a golden match for photometric BHB studies providing, at the same time, unprecedented depth as well as purity and completeness. 

\section{Results: Investigating the halo with BHB and BS stars}\label{results}

As mentioned above, BHB stars are standard candles, which means that distances can be computed for them using their apparent magnitude. We again here follow \citet{Deason11}, who define

\begin{equation}\label{eq:distBHB}
\begin{split}
M_{g,\rm BHB} =\ & 0.434 - 0.169(\gmr)+2.319(\gmr)^{2} \\
& + 20.449(\gmr)^{3} + 94.517(\gmr)^{4}.
\end{split}
\end{equation}

Additionally, the BS stars can be used to trace substructures \citep[e.g.,][]{Thomas18b}. For the BS stars, we use the conversion to absolute magnitude (and hence distance) as given by \citet{Navarrete18}:

\begin{equation}\label{eq:distBS}
M_{g, \rm BS} = 2.2 + 4.557 (g_{0}-r_{0}) - 0.45\FeH.
\end{equation}

The BS stars are less accurate standard candles and for them the distance vs. absolute magnitude relation will have a higher dispersion than for BHB stars. Additionally, the fact that they are intrinsically fainter means that they do not trace as far out and will suffer at closer distances from larger photometric uncertainties. In the absence of any robust measurement of the metallicity, we use a metallicity of [Fe/H] = $-1.0$ in accordance with our results in Figure \ref{fig:separateBSBHB}. This lower metallicity will be a more accurate estimate for the BS stars further out in the halo than for those in the disk. 

We note that the absolute magnitude for a typical A star will differ by about 1.5 magnitudes if it is a BHB or a BS star. As pointed out by \citet{Lancaster19}, this means that the proper motions of the stars (if available) as measured by Gaia DR2 can provide an extra means to clean the sample. A BS star that is misclassified as a BHB star will as a result be placed much further and its tangential velocity calculated will be larger based on the same proper motion. In this work, we adopt as an extra criterion that if the tangential velocity calculated with the corresponding BHB distance for a star is above 500 km s$^{-1}$, a limit of 0.4 is adopted for the BHB probability. If the probability that the star is a BHB star was above this value before, it is lowered to 0.4 and the probability that the star is a BS star is raised to 0.6. This is only applied if the relative proper motion uncertainty is smaller than 0.2. The effect of this extra cleaning is small, it does not affect any stars in the overlapping spectroscopic sample and less than 1\% of the stars in our total footprint ($\sim 2.5\%$ of the stars with P $> 0.5$ of being a BHB star). We furthermore use the same colour and quality cuts as described in Section \ref{sec:comp}, but with no strict cut on $CaHK$ magnitude uncertainty (although in practice very few stars have uncertainties larger than the value of 0.1 that was used as a limit before). We verify with the second data release from Gaia \citep{Brown18} that the distances we obtain for the samples are in agreement with their independently measured parallax distances. As expected, however, most of the sample is too faint to have accurate parallax measurements. But for stars with relative parallax uncertainty smaller than 0.2, all distances are compatible with the Gaia measurements.

\subsection{The profile of the Galactic halo}

Figure \ref{fig:disthist} shows the histogram of heliocentric distances for three different populations: the BHB population as identified in Section \ref{SDSSalone} from only SDSS photometry and the BHB population as well as the BS population from combining \pris and SDSS photometry as described in Section \ref{SDSSandPris}. All stars selected here have a probability of $> 0.8$ to belong to their class, which means that the completeness of the sample will sharply decline between $\sim$50 and $\sim100$ kpc (see Figure \ref{fig:evmag}). We find an enhancement of BHB stars at all distances in the \pris + SDSS sample, most likely because of the improvement in completeness. Most strikingly, there is a clear extra overdensity at 45--50 kpc and $200<\alpha<220^\mathrm{o}$ in the \pris + SDSS BHB sample. This feature is again seen clearly when the distances are shown as a function of right ascension in Figure \ref{fig:skydist}. Here, stars with a probability $> 0.5$ of belonging to a class are shown in each panel (in cyan, 2720 stars), as well as those that have probability $> 0.8$ (in black, 2311 stars). As expected, the samples start to deviate at larger distances, where the photometric uncertainties are larger, and start to match the separation between the ridgelines in Figure \ref{fig:separateBSBHB}. The larger separation between the ridgelines in the \pris + SDSS sample lead to this effect being smaller at similar distances than in the SDSS only sample. 

The feature at $\sim50$ kpc can be easily identified as the leading arm of the Sagittarius stream, which indeed is expected to cross the \pris footprint at these distances. Another overdensity, seen at $\alpha\sim120^\mathrm{o}$ and $\sim90$ kpc represents the trailing arm of the same disrupting dwarf galaxy. Figure \ref{fig:Sagfootprint} presents an overview of the \pris footprint to date in the coordinates of the Sagittarius orbital plane (see, e.g., \citealt{Belokurov14} for its definition), with overplotted the \citet{Law10} model of the Sagittarius stream and its predicted distances, matching well the observed relation with the \pris footprint in Figure \ref{fig:skydist}. Interestingly, the stream is not clearly traced in the BS star sample, but it does stand out clearly in the BHB Pristine + SDSS sample, and more clearly than the BHB stars selected from SDSS photometry alone. 

Because of the overwhelming presence of the Sagittarius stream in our footprint, the \pris dataset is currently not well suited to study the halo profile of the smooth(er) halo component. There has been a large body of literature already on this subject, utilising more extended and deeper datasets, including BHB stars, K-giants, or RR Lyrae \citep[e.g.,][]{Xue15, Xu18,Hernitschek18,Thomas18a,Fukushima18}. However, given our improved statistics on completeness and purity, it is interesting to compare our findings of the halo density profile with other work. In Figure \ref{fig:profile}, we summarise the density profile of our \pris + SDSS BHB distribution over the full footprint (filled circles) and for just the part of the footprint that is more than 30 degrees away from the Sagittarius orbital plane (a bit less than 20\% of our A stars, shown as asterisks). In the full sample, clearly the influence of the Sagittarius stream at $\sim$50 kpc can be detected. To address the issue of completeness, we have counted the number density of BHB stars in each distance bin in three different ways: (i) by counting the stars for which $P({\rm BHB}~|~uCaHKgr) > 0.5$; (ii) by counting only those for which $P({\rm BHB}~|~uCaHKgr) > 0.8$; (iii) by using all stars that pass the quality cuts and multiplying them with their value of $P({\rm BHB}~|~uCaHKgr)$. These three different choices are plotted with three different colours in Figure \ref{fig:profile}, but it is clear that the exact choice does not affect the overall results except in the outermost distance bins.

When fitted by a single power-law using the least-square polynomial fit routine \texttt{poly\_fit} in IDL, the density distribution favours a slope of $-3.5 \pm 0.1$ for the full sample and a slope of $-4.0 \pm 0.1$ for the sample off the Sagittarius plane, which is in good agreement with recent results from either RR Lyrae \citep[$-4.4\pm0.1$][]{Hernitschek18}, or BHB stars \citep[measurements ranging from $-3.2$ to $-4.0$][]{Thomas18a, Deason18, Fukushima18}, although the RR Lyrae favour a slightly steeper slope. Results from \citet{Xu18} using a K-giant sample also favour a steeper slope of $-5.0 \pm 0.6$ and additionally a significant flattening in the inner halo (although their preferred profile becomes nearly spherical beyond 30 kpc). Note that we do not attempt to additionally fit for the flattening of the halo. We see no strong evidence for a flattening of the slope in the inner region and/or a steeper slope at larger distances, which was reported by \citet{Deason11} and \citet{Deason14} who measured slopes of $-2.3$ up to 27 kpc, a slope of $-4.6$ beyond that and subsequently a strong steepening to $-6$ or even $-10$ beyond 50 kpc. When we allow for a broken power-law fit, using the fitting method outlined in \citet{Thomas18a}, the preferred break radius is at $\sim$73 kpc for the P $> 0.5$ sample, both for the full sample and the sample off the Sagittarius plane, after which the slope steepens. However, it is difficult to rule out that this is not to be attributed to the declining completeness and purity in our samples at these distances (as shown in Figure \ref{fig:evmag}). Deeper photometry is required to investigate this further.

\subsection{Clumpiness in the Galactic halo}

Figures \ref{fig:disthist} and \ref{fig:skydist} show that there is significant deviation from smoothness in the Galactic halo between the different selected samples. Here, we aim to quantify this signal using a pair counting method to measure the amount of clustering in each sample on different scales \citep[we refer the reader to][and Youakim et al., in prep., for a more elaborate analysis using a similar method with RR Lyrae and FGK stars, respectively]{Lancaster2018}. 

\begin{figure}
\begin{center}
\includegraphics[width=\columnwidth, angle=0]{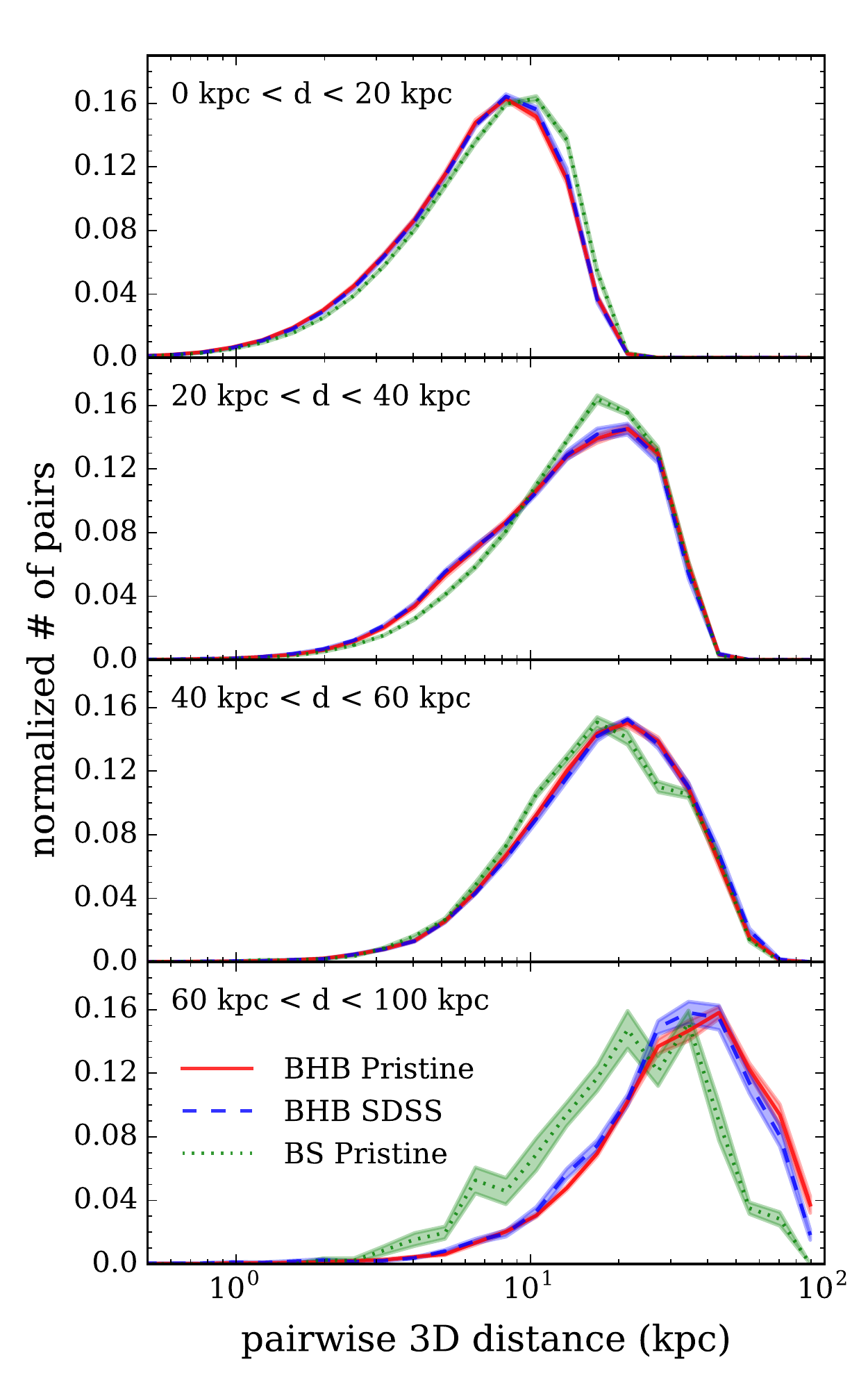}
\caption{Pairwise 3D clustering for the BHB population as identified in Section \ref{SDSSalone} from SDSS photometry alone (blue) and the BHB population as well as the BS population from combining \pris and SDSS photometry as described in Section \ref{SDSSandPris} (red and green respectively). \label{fig:clust}}
\end{center}
\end{figure}

In Figure \ref{fig:clust}, we show the results of a clustering analysis of the BHB samples selected with \pris + SDSS against the sample selected with SDSS alone. Additionally, we include the BS star sample from the \pris + SDSS selection. We select any members that have a probability $>0.5$ to belong to their class. To obtain the clustering signal, 3-dimensional separations are calculated within a given distance bin. The curves shown in Figure \ref{fig:clust} can be interpreted as the relative amount of clustering on a given scale and are normalised to emphasise their different shapes, rather than different numbers of stars in each class. We limit this analysis to the largest contiguous region of the footprint, for $180<\alpha<260^\mathrm{o}$, so as to avoid pair count contributions at large separations from region to region. The error bars in Figure \ref{fig:clust} represent the standard deviation of 10 bootstrapping iterations of counting these pairs, each time taking a subsample of 80\% of the total samples. Individual uncertainties on distances are not taken into account here, but we have tested that distance uncertainties of the order of 10\% do not change the result. Typically, to measure clustering one would not only do pair counts, but also compare these to pair counts of a suitable random distribution. Given that this is a relative comparison between samples using the same underlying data quality cuts and in the same footprint, we do not have to rely on the creation of a random sample since it would be the same for all of the samples. Furthermore, creating such a (smooth) random sample is non trivial and would rely heavily on certain assumptions on the selection functions of the surveys and the underlying density gradients of the various Galactic components.

\begin{figure*}
\begin{center}
\includegraphics[width=1\linewidth]{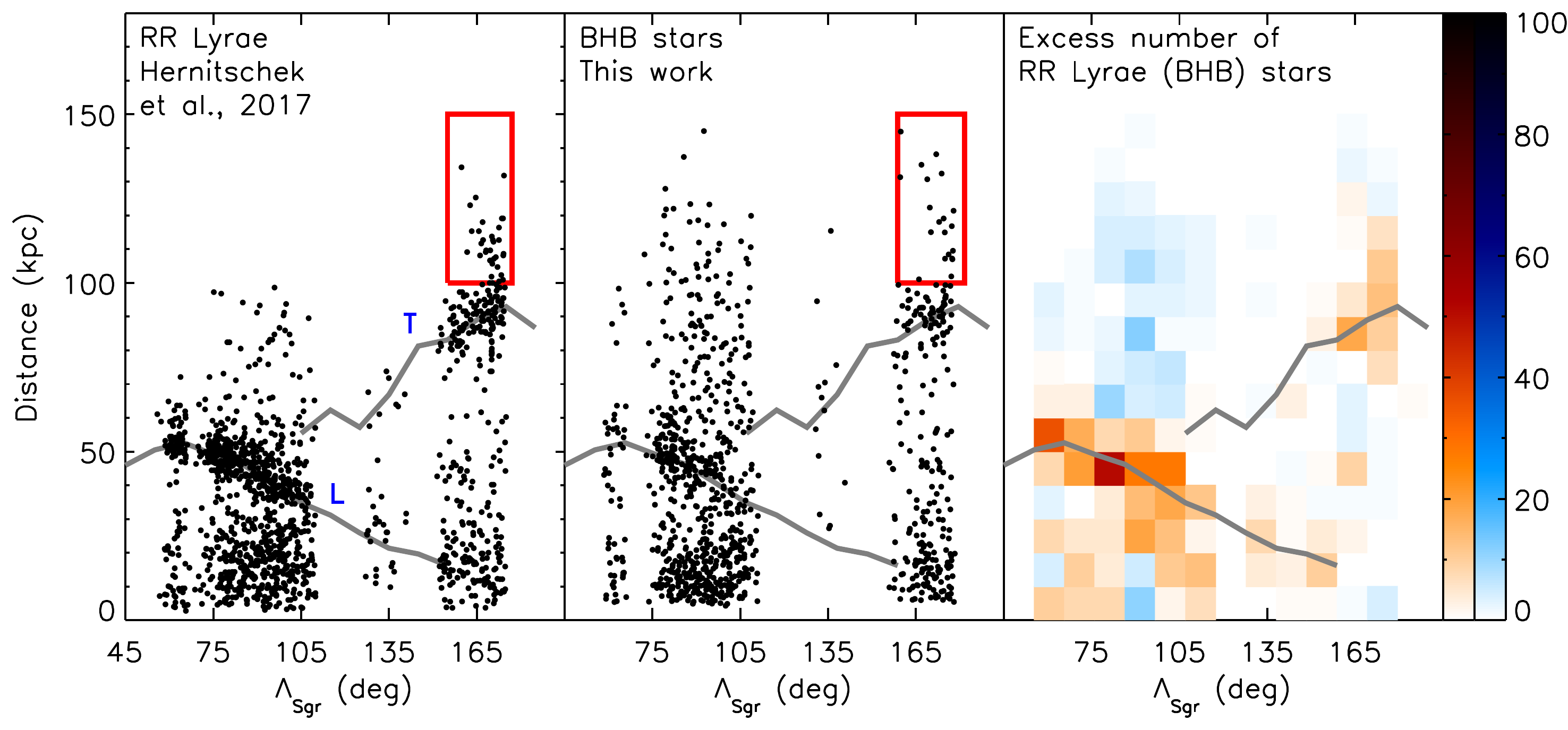}
\caption{RR Lyrae (left panel, taken from \citet{Hernitschek17}) and BHB stars (second panel, this work, selected to have $P({\rm BHB}~|~uCaHKgr) > 0.5$) as a function of longitude along the Sagittarius orbital plane and heliocentric distance. Grey lines guide the eye along Sagittarius stream components as defined by the RR Lyrae measured by \citet{Hernitschek17}. Blue letters `L' and `T' indicate the leading and trailing arm respectively. Red boxes illustrate the distant ``spur'' feature discovered in the same RR Lyrae sample \citep{Sesar17b}. The third panel illustrates the difference in both populations by either counting the excess of RR Lyrae stars (red colour bar), or BHB stars (blue colour bar). \label{fig:spur}}
\end{center}
\end{figure*}

The results in Figure \ref{fig:clust} are divided up in four different distance ranges. We see how in the first distance bin, up to 20 kpc, the clustering behaviour is very similar in all three samples. At 20-40 kpc, there is a feature at a scale of 5 kpc for the BHB samples indicating a higher degree of clustering at small scales compared to the BS sample. At the peak of the distribution at a scale of 20 kpc, there is relatively less clustering in the BHB samples when compared to the BS stars. This is indicative that the BHB samples are better tracing the strong feature of the Sagittarius stream at these distances, which is pronounced at the edge of the distance bin, but is largely under-dense in the middle. By comparison, the BS sample is more uniformly distributed. In general, BS stars do not trace substructures as well as BHB stars, due to the larger uncertainties in their distance determinations. They also trace a younger stellar population than the BHB stars do. In the 40-60 kpc bin, all three samples show a clustering signature with the same peak, however, the BS sample shows a decrease in clustering at slightly larger scales compared to the BHB samples. Referring back to Figure \ref{fig:skydist}, the BHB signal is again explained by a strong Sagittarius feature, whereas the BS signal comes from a different feature, characterised by a larger density of stars in the low RA half of the footprint compared to the high RA half. This either indicates some other structure that is traced by the BS stars but not the BHB stars, possibly characterised by a younger stellar population, or that there is some systematic selection effect in that region for the BS sample. Finally, in the largest distance bin (60--100 kpc) a hint of an effect in the clustering signal of our improved selection technique for BHB stars with \pris + SDSS can be seen. Here, the sample selected on SDSS alone shows a larger clustering signal at smaller scales of roughly 40 kpc. By comparison, the \pris + SDSS sample shows stronger clustering at larger separations. These differences are not very significant however and need to be investigated further with larger samples and more accurate photometry at these large distances. If confirmed, the differences can be related to two things, firstly a misclassification of BS stars that should appear in the more nearby distance bins but are instead pushed into the larger distance bin in the SDSS only sample, and secondly that there is a larger number of stars at the furthest distance bin in the \pris + SDSS sample (see Figure \ref{fig:disthist}). Together, these result in a broader clustering signal which extends to larger scales for the \pris + SDSS sample.

\subsection{Tracing the Sagittarius stream further out}

In Figure \ref{fig:spur}, we show how the Sagittarius stream is traced in the current \pris survey footprint when compared with the RR Lyrae stars presented in \citet{Hernitschek18} across the same footprint (left panel). The selection procedure for the RR Lyrae sample is described in much detail in \citet{Sesar17a}. Besides reaching distances of $>$120 kpc with a very high distance precision (3\%), this sample is both very pure (90\%) and reaches a high level of completeness (80\% at 80 kpc). For the BHB sample, we use a magnitude limit for our sample to 21.5 for this analysis, allowing BHB stars in principle to be traced out to 170 kpc for the hottest stars, although probably with significant contamination. A BHB criterion with a probability of $> 0.5$ is adopted. Grey solid lines guide the eye along the Sagittarius main stream features as defined for RR Lyrae by \citet{Hernitschek18}. We clearly see the same structures appear in our map of BHB stars, also illustrating that the calculated distances match very well the precise distances to RR Lyrae stars. 

As both populations of stars have a high completeness and purity up to 50 kpc, we can learn about the composition of these different features through the observed ratio of RR Lyrae to BHB stars. The right panels of Figure \ref{fig:spur} show this comparison. It is clear that in the Sagittarius features the signal is largely dominated by the RR Lyrae. Both RR Lyrae and BHB stars are thought to trace old stellar populations. One possible interpretation of this difference is that the outer halo in the \pris survey footprint is made up from progenitor galaxies with different dominant stellar populations than the Sagittarius dwarf galaxy, for instance being smaller and more metal-poor. This would explain why the Sagittarius stream features are more dominated by RR Lyrae. Note that this could also explain the slightly steeper slope found in the halo using RR Lyrae as tracers \citep{Hernitschek18}, when compared to BHB stars \citep{Thomas18a, Deason18, Fukushima18}. There is also a larger number of BHB stars found in the more distant regimes (most clearly visible in the region $75 < \Lambda_{\rm Sgr} < 105^\mathrm{o}$, between 100 and 120 kpc), but this could well be an effect of the declining performance in completeness and purity of the BHB samples.

Many authors have tried to model the Sagittarius streams, most often in $\Lambda$ Cold Dark Matter ($\Lambda$CDM) cosmologies, but also some modelling has been done using alternative cosmological frameworks such as MOND \citep{Thomas17}. No model to date has however been able to reproduce all features simultaneously \citep[see the discussion in][]{Fardal19}. This might point to a more complicated halo dark matter profile than is usually assumed. Additionally, in the mapping of its orbit, the internal dynamics of the dwarf galaxy and the influence of other large substructures, such as the Large Magellanic Cloud cannot always be ignored \citep{Penarrubia10,VeraCiro13} and the Sagittarius dwarf galaxy itself might perturb and alter the Milky Way potential significantly \citep[e.g.,][]{Purcell11,Gomez15}. 

Recent observations put the apocentre distances of the leading and trailing stream at very different distances, roughly a factor two apart at $\sim$50 and $\sim$100 kpc \citep{Belokurov14,Sesar17b,Hernitschek18}. Moreover, the observational data clearly hints at an extension of Sagittarius' features even beyond the furthest apocentre at $\sim$100 kpc. The extent and character of the ``spur'' feature, seen in the RR Lyrae about 30 kpc beyond the apocentre of the trailing arm itself \citep{Sesar17b} will provide some further crucial progress on understanding the Sagittarius stream.

While many of the modelling efforts fail to reproduce this behaviour \citep[for instance, the model of][puts the furthest apocentre at $\sim65$ kpc]{Law10}, two recent modelling efforts, by \citet{Dierickx17} and \citet{Fardal19}, both reproduce the observed apocentre distances of the leading and trailing stream as well as a clear ``spur''-like feature. \citet{Fardal19} interpret this feature as trailing debris from two different pericentric passages, whereby the outer material actually represents a more recent passage. In their modelling the debris reaches about 140 kpc in heliocentric distance. \citet{Dierickx17} instead present a model where this feature reaches a heliocentric distance of 250 kpc, including a prediction for the velocity profile of these stars. 

We are fortunate to find this intriguing ``spur'' feature in our footprint (marked with a red rectangle in Figure \ref{fig:spur}). Tentatively, we can follow this structure further out using the BHB stars instead of the RR Lyrae, but unfortunately not far enough out yet to favour the model of either \citet{Dierickx17}, or \citet{Fardal19}. Additionally, it is uncertain how much trust we can put in the discovery of this feature at these large distances. The completeness of the BHB sample should still be around 40\%, according to our modelling shown in Figure \ref{fig:evmag}, but the purity of the sample at these distances is unknown and could well be very poor. We will pursue a further characterisation of this feature in future work by using better quality photometry and targeted spectroscopic follow-up of candidate BHB stars at these distances.

\section{Conclusions}\label{conc}

In this work, we demonstrate how the narrow-band filter designed for the \pris survey for its metallicity sensitivity additionally provides an excellent discriminating power to separate standard candle BHB stars from the contaminating population of intrinsically fainter BS stars (for which distance determinations are less accurate). Using this narrow-band filter in combination with SDSS $u$-band information we improve the purity of the BHB-star selection from 84 to 93\% while at the same time increasing the completeness of the sample from 67 to 91\%. 

Using this unprecedented clean and complete sample of BHB and BS stars we trace the outer halo of the Milky Way. Their distance profile follows a power-law with an almost constant negative slope of 3.5--4.0, depending on whether we select regions on or off the main Sagittarius stream. We investigate how our cleaner and more complete selection affects a quantification of the clumpiness in this part of the halo in Figure \ref{fig:clust}, hinting at a less tightly clumped signal at large distances than a sample that is selected solely using broad-band colours. Our mapping of the Sagittarius stream is very much in agreement with results using RR Lyrae \citep[][]{Hernitschek18}, as shown in Figure \ref{fig:spur}, illustrating however also that the BHB stars are dominating the smoother halo component while the Sagittarius stream is dominated by RR Lyrae. This likely reflects a different parent stellar population for these halo features, for instance in metallicity.
 
Additionally, we have laid some groundwork for subsequent follow-up of the recently discovered ``spur'' feature, which represents the furthest discovered (likely) feature of the Sagittarius stream. Tentatively, we trace this feature to larger distances; this, however, needs to be confirmed with better quality data. Spectroscopic follow-up could provide additional proof that these stars are indeed BHB stars as well as radial velocities of the stars in this feature, providing another avenue to constrain its origin.

\section*{Acknowledgements}
We gratefully thank the CFHT staff for performing the observations in queue mode, for their reactivity in adapting the schedule, and for answering our questions during the data-reduction process. We thank Nina Hernitschek for granting us access to the catalogue of Pan-STARRS1 variability catalogue. 

ES, KY, and AA gratefully acknowledge funding by the Emmy Noether program from the Deutsche Forschungsgemeinschaft (DFG). This work has been published under the framework of the IdEx Unistra and benefits from a funding from the state managed by the French National Research Agency as part of the investments for the future program. NFM, RI, and NL gratefully acknowledge support from the French National Research Agency (ANR) funded project ``Pristine'' (ANR-18-CE31-0017) along with funding from CNRS/INSU through the Programme National Galaxies et Cosmologie and through the CNRS grant PICS07708. ES, KY, NM, AA, JIGH, and NL benefited from the International Space Science Institute (ISSI) in Bern, CH, thanks to the funding of the Teams ``The Formation and Evolution of the Galactic Halo'' and ``Pristine''. JIGH acknowledges financial support from the Spanish Ministry project MINECO AYA2017-86389-P, and from the Spanish MINECO under the 2013 Ram\'on y Cajal program MINECO RYC-2013-14875.

Based on observations obtained with MegaPrime/MegaCam, a joint project of CFHT and CEA/DAPNIA, at the Canada-France-Hawaii Telescope (CFHT) which is operated by the National Research Council (NRC) of Canada, the Institut National des Sciences de l'Univers of the Centre National de la Recherche Scientifique of France, and the University of Hawaii. SDSS-IV is managed by the Astrophysical Research Consortium for the 

The Pan-STARRS1 Surveys (PS1) have been made possible through contributions of the Institute for Astronomy, the University of Hawaii, the Pan-STARRS Project Office, the Max-Planck Society and its participating institutes, the Max Planck Institute for Astronomy, Heidelberg and the Max Planck Institute for Extraterrestrial Physics, Garching, The Johns Hopkins University, Durham University, the University of Edinburgh, Queen's University Belfast, the Harvard-Smithsonian Center for Astrophysics, the Las Cumbres Observatory Global Telescope Network Incorporated, the National Central University of Taiwan, the Space Telescope Science Institute, the National Aeronautics and Space Administration under Grant No. NNX08AR22G issued through the Planetary Science Division of the NASA Science Mission Directorate, the National Science Foundation under Grant No. AST-1238877, the University of Maryland, and Eotvos Lorand University (ELTE).

Funding for the Sloan Digital Sky Survey IV has been provided by the Alfred P. Sloan Foundation, the U.S. Department of Energy Office of Science, and the Participating Institutions. SDSS-IV acknowledges support and resources from the Center for High-Performance Computing at the University of Utah. The SDSS web site is www.sdss.org. SDSS-IV is managed by the Astrophysical Research Consortium for the Participating Institutions of the SDSS Collaboration including the Brazilian Participation Group, the Carnegie Institution for Science, Carnegie Mellon University, the Chilean Participation Group, the French Participation Group, Harvard-Smithsonian Center for Astrophysics, Instituto de Astrof\'isica de Canarias, The Johns Hopkins University, Kavli Institute for the Physics and Mathematics of the Universe (IPMU) / University of Tokyo, Lawrence Berkeley National Laboratory, Leibniz Institut f\"ur Astrophysik Potsdam (AIP), Max-Planck-Institut f\"ur Astronomie (MPIA Heidelberg), Max-Planck-Institut f\"ur Astrophysik (MPA Garching), Max-Planck-Institut f\"ur Extraterrestrische Physik (MPE), National Astronomical Observatories of China, New Mexico State University, New York University, University of Notre Dame, Observat\'ario Nacional / MCTI, The Ohio State University, Pennsylvania State University, Shanghai Astronomical Observatory, United Kingdom Participation Group,Universidad Nacional Aut\'onoma de M\'exico, University of Arizona, University of Colorado Boulder, University of Oxford, University of Portsmouth, University of Utah, University of Virginia, University of Washington, University of Wisconsin, Vanderbilt University, and Yale University.




\bibliographystyle{mnras}
\bibliography{references} 






\bsp	
\label{lastpage}
\end{document}